\documentclass[aps,prl,twocolumn,showpacs,superscriptaddress,showkeys,bm,amsmath,amssymb]{revtex4-1}
\pdfoutput=1

\usepackage{fixmath}
\usepackage{enumitem}
\usepackage[dvipsnames]{xcolor}
\usepackage{amsmath}
\usepackage{mathtools}
\usepackage[utf8]{inputenc}
\usepackage[english]{babel}
\usepackage{amsmath}
\usepackage{graphicx}
\usepackage{epsfig}
\usepackage{slashed}
\usepackage{amssymb}
\usepackage{color}
\usepackage{bm}
\usepackage{booktabs}
\usepackage{float}
\usepackage{array}
\usepackage{tabulary}
\usepackage{makecell}
\usepackage{enumitem}
\usepackage{tabularx} 

\usepackage{multirow}
\usepackage[normalem]{ulem}
\definecolor{MyDarkBlue}{rgb}{0.1, 0.1, 0.8} 
\definecolor{SBlue}{rgb}{0.2, 0.4, 0.7} 
\definecolor{MyLightBlue}{rgb}{0.22,0.51,0.9}
\definecolor{MyGreen}{rgb}{0.0, 0.5, 0.0}
\definecolor{BrickRed}{rgb}{0.8, 0.25, 0.33}
\RequirePackage{hyperref}
\hypersetup{colorlinks, citecolor=SBlue,linkcolor=MyDarkBlue, urlcolor=PineGreen}
\makeatletter
\usepackage{amssymb}
\usepackage{pifont}

\begin{document}

\title{\Large 
Search for Leptophilic Dark Matter at the LHeC
}

\author{Guo-yuan Huang}
\email[E-mail: ]{guoyuan.huang@mpi-hd.mpg.de}
\affiliation{Max-Planck-Institut f{\"u}r Kernphysik, Saupfercheckweg 1, 69117 Heidelberg, Germany}

\author{Sudip Jana}
\email[E-mail: ]{sudip.jana@mpi-hd.mpg.de}
\affiliation{Max-Planck-Institut f{\"u}r Kernphysik, Saupfercheckweg 1, 69117 Heidelberg, Germany}

\author{Álvaro S.\ de Jesus}
\email[E-mail: ]{alvarosdj@ufrn.edu.br}
\affiliation{International Institute of Physics, Universidade Federal do Rio Grande do Norte,
	Campus Universitario, Lagoa Nova, Natal-RN 59078-970, Brazil}
\affiliation{Departamento de Fisica, Universidade Federal do Rio Grande do Norte, 59078-970, Natal, RN, Brasil}

\author{Farinaldo S.\ Queiroz}
\email[E-mail: ]{farinaldo.queiroz@ufrn.br}
\affiliation{International Institute of Physics, Universidade Federal do Rio Grande do Norte,
Campus Universitario, Lagoa Nova, Natal-RN 59078-970, Brazil}
\affiliation{Departamento de Fisica, Universidade Federal do Rio Grande do Norte, 59078-970, Natal, RN, Brasil}
\affiliation{Millennium Institute for SubAtomic Physics at the High-energy frontIeR, SAPHIR, Chile}

\author{Werner Rodejohann}
\email[E-mail: ]{werner.rodejohann@mpi-hd.mpg.de}
\affiliation{Max-Planck-Institut f{\"u}r Kernphysik, Saupfercheckweg 1, 69117 Heidelberg, Germany}

\begin{abstract}
\noindent 
The Large Hadron electron Collider (LHeC) has been designed to push the field of deep inelastic scattering to the high energy and intensity frontier using an intense electron beam with a proton beam from the High Luminosity–Large Hadron Collider. However, LHeC is also a great laboratory for new physics. In this work, we propose a search for dark matter that couples with leptons. This may yield $ej + \slashed{E}$ and $\mu j + \slashed{E}$ signals that can be potentially observed through simple missing-energy cuts that suppress the Standard Model background. Considering direct dark matter detection and LHC constraints, we show that LHeC can indeed discover a weak scale dark matter fermion for masses up to  $350$~GeV, which reproduces the correct relic density, and has interesting implications for lepton flavor violation.
\end{abstract}

\maketitle

\section{Introduction}
\noindent

There is an extensive program dedicated to dark matter searches worldwide. Dark matter particles might leave imprints at direct and indirect detection experiments. However, due to the nature of these searches which are subject to nuclear and astrophysical uncertainties, respectively, it is desirable that one could also observe a dark matter signal at colliders or accelerators. The properties of the dark matter particle dictate the type of signals one may observe. Several collider searches have been conducted at the Large Hadron Collider (LHC) and other accelerators, but thus far no positive result signal has been found. 

The LHC is an excellent probe when dark matter particles interact with quarks. Vector boson fusion production of dark matter yields interesting but less restrictive  constraints \cite{Abercrombie:2015wmb}. If dark matter features leptonic interactions, obviously an $e^+ e^-$ collider \cite{Dreiner:2012xm} or an LHeC that features an electron beam allowing synchronous operation of $ep$ with $pp$ collisions at the LHC  \cite{LHeCStudyGroup:2012zhm} would be helpful. Future colliders such as the Future Circular Collider \cite{FCC:2018evy} and the International Linear Collider \cite{Baer:2013cma} are very long-term proposals, and LHeC can be seen as a very important step towards this major new facilities in terms of physics as well as technology. That said, could LHeC probe weak scale dark matter? As usual, dark matter particles are missing energy at colliders as they simply leave no imprint at the detectors, and a necessary visible counterpart is required as a trigger.

If dark matter can be produced via $s$-channel diagrams, one can take advantage of resonance production of the mediators and quickly conclude that LHeC cannot be regarded as the best probe for such dark matter particles. However, if dark matter is produced via leptonic processes, LHeC stands out. We will present here the LHeC sensitivity to dark matter particles that have couplings to leptons. Without loss of generality, we assume the following Yukawa interaction 
\begin{equation}
\label{eq:1}
\mathcal{L}_Y \supset  y^{}_{N\ell} \eta \overline{N} L_{\ell} + {\rm h.c.} 
\end{equation}
Here $N$ is a Majorana singlet fermion and our dark matter particle, $\eta$ is a scalar SM doublet, $L$ is a lepton doublet, and $\ell=e,\mu$. This Yukawa interaction is often present in neutrino mass models \cite{Ma:2006km}, but in our work it is treated more generally. A $Z_{2}$ symmetry has been imposed to the new particles to forbid the Majorana fermion to couple with other SM states, warranting its stability, as long as its mass is lighter than that of the inert doublet. The doublet is inert either by symmetry requirements or by theoretical constructions. The dark matter abundance of the Majorana fermion occurs solely through $t$-channel processes into SM leptons in the standard freeze-out regime.  There are no dark matter direct detection signals at tree level. At one-loop level, one can potentially induce dark matter-nucleus scattering with charge-charge, dipole-charge, and dipole-dipole interactions \cite{Schmidt:2012yg}. After investigating the collider phenomenology, we put our findings into perspective with lepton flavor violation, namely $\mu \rightarrow e\gamma$ \cite{Toma:2013zsa,Lindner:2016bgg}, and discuss the importance of direct dark matter detection. 

In summary, Eq.\ (\ref{eq:1}) is commonly presented in a multitude of models, and has implications for dark matter, lepton flavor violation as well as neutrino mass models. In what follows, we will assess the LHeC potential to discover such dark matter particle to solidly conclude that LHeC offers a unique opportunity to discover a weak scale dark matter particle.
\newpage


\section{Signature of Weak Scale Dark Matter at the LHeC}
\noindent

As explained above, there are benefits of the electron-proton collider: a relatively clean environment for new physics searches, and the possibility to probe weak scale dark matter that couples  to leptons. 

\begin{figure}[t!]
	\begin{center}
		\includegraphics[width=0.5\textwidth]{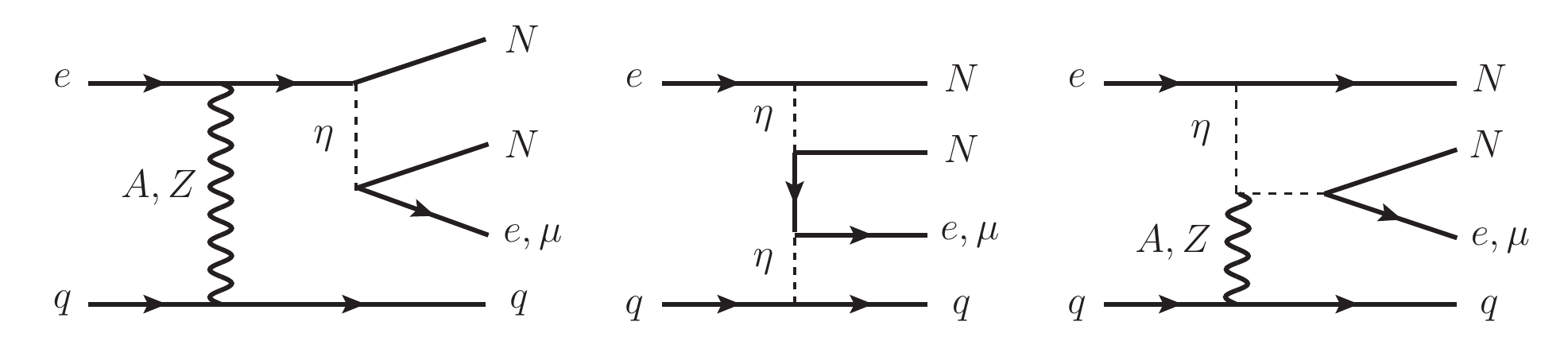}
	\end{center}
	\vspace{-0.3cm}
	\caption{Feynman diagrams contributing to the typical events $ej + \slashed{E}$ and $\mu j + \slashed{E}$.}
	\label{fig:FD}
\end{figure}

\subsection{Relevant Event Topologies}
\noindent
The relevant Feynman diagrams are shown in Fig.~\ref{fig:FD}. We consider each channel at a time. We find that the effective Lagrangian can induce the following event topologies from  electron-proton scattering at LHeC:
\begin{itemize}
	\item $ej + \slashed{E}$: The contributed signal process reads $e p\to ejNN$, where the dark matter $N$ generates missing energy. The SM backgrounds for this channel include the following: (i) $e p \to e j X$ with $X$ ($\gamma$, $G$, etc.); (ii) $ep \to ej \nu\nu$; (iii) $e p \to e j \nu X$ ($X=\ell$); (iv) $ep \to ejXX$ ($X= \gamma,G,q, \ell$). 
	Here and in the following, $X$ denotes an in principle  visible particle that however escapes the angular acceptance of the detector. 
	Note that we use $\nu$ in the final state to denote both neutrino and antineutrino for all possible flavors.
	In this analysis, we do not distinguish  light quarks and gluons in the final state.
	Processes with even higher orders, e.g., $e p \to e j \nu \nu X$, are neglected. 
	\item $\mu j + \slashed{E}$: The final state with muons has fewer SM backgrounds because of the suppression of lepton flavor violation. In comparison to the $e j + \slashed{E}$ final state, we identify three types of SM backgrounds: (i) $e p \to \mu j \nu \nu $; (ii) $e p \to \mu j \nu X$ ($X=e^-,\mu^+$); (iii) $e p \to \mu j X X$ ($X=e^-,\mu^+$).
	\item Mono $j+\slashed{E}$: The mono-jet topology corresponds to the process $e q \to   j\nu N N$. There are the following SM backgrounds: (i) $e q \to j \nu(X)$; (ii) $e q \to j \nu X$; (iii) $e q \to j \nu(X) \nu(X) \nu(X)$.
	The leading-order process $e q \to j \nu$ will be an overwhelmingly large background and cannot be reduced by imposing  missing energy cuts. 
	Thus, we will focus on the previous two topologies in our analysis.
\end{itemize}
Some of those backgrounds have a similar diagrammatic pattern as the signal channel, hence  irreducible. However,  some of them can be reduced by choosing proper cuts which we will describe later.

\begin{table*}[t!] 
	\centering
	\begin{tabular}{|c| c|  c|c|c  | c|} 
		\hline
		Event Topology & Signal $\sigma$ (fb) & \multicolumn{4}{|l|}{  \hspace{2.5cm} Background $\sigma$ (fb)} \\ 
		\hline
		\multirow{2}{*}{$ej + \slashed{E}$} & \multirow{2}{*}{24.4 (16.6)}  & $e j X$ & $e j \nu \nu $ & $e j \nu X$ & $e j  X X$ \\ \cline{3-6}
		& & $1.83\times 10^5$ ($0$)  & $92.9$ ($33.1$) & $73.1$ ($2.20$) & $5.20 \times 10^5$ ($0$) \\ \hline
		\multirow{2}{*}{$\mu j + \slashed{E}$} & \multirow{2}{*}{24.5}  & $\mu j X$ & $\mu j \nu \nu $ & $\mu j \nu X$ & $\mu j  X X$ \\ \cline{3-6}
		& & 0 & 63.8 & 3.01 & 0.140 \\ \hline
	\end{tabular}
	\caption{The cross sections for two event topologies $ej + \slashed{E}$ and $\mu j + \slashed{E}$, assuming 60 GeV electrons colliding with 7 TeV protons. For the signal cross section, the model parameters have been taken as $M_\eta=300~{\rm GeV}$, $M^{}_{N}=100~{\rm GeV}$ and $y^{}_{Ne}=y^{}_{N\mu }=1$. For $ej + \slashed{E}$, the cross section is given with the cut $\slashed{p}^{}_{\rm T} \gtrsim 10~{\rm GeV} $ (or $\slashed{p}^{}_{\rm T} \gtrsim 100~{\rm GeV}$ for the value in parentheses). The label `$X$' in the table denotes  visible particles which have however escaped the detector angular acceptance, e.g., $\gamma$, $G$, $q$ and  $\ell$. }
	\label{table:xsec}
\end{table*}

\subsection{Sensitivity to the Parameter Space}
\noindent
The sensitivity is obtained by a simple count analysis of total events without involving  detailed event distributions.
Hence, the present analysis represents a conservative result.
We adopt the following chi-square to estimate the sensitivity:
\begin{eqnarray} \label{eq:chi2}
\chi^2(M^{}_{N}, M^{}_{\eta}, y^{}_{N}) = \frac{N^2_{\rm sig}}{N^{}_{\rm sig}+N^{}_{\rm bkg}}\;,
\end{eqnarray}
where $N^{}_{\rm sig}$ and $N^{}_{\rm bkg}$ are the total signal and background event numbers, respectively.
The following cut will be implemented at the parton level: $p^{}_{\rm T}(j/b) > 20~{\rm GeV}$, $p^{}_{\rm T}(\ell) > 10~{\rm GeV}$, and the pseudorapidity $|\eta(j/b/\ell)|<5$~\cite{Cottin:2021tfo} (see also \cite{Lindner:2016lxq}).

\begin{figure}[t!]
	\begin{center}
		\includegraphics[width=0.3\textwidth]{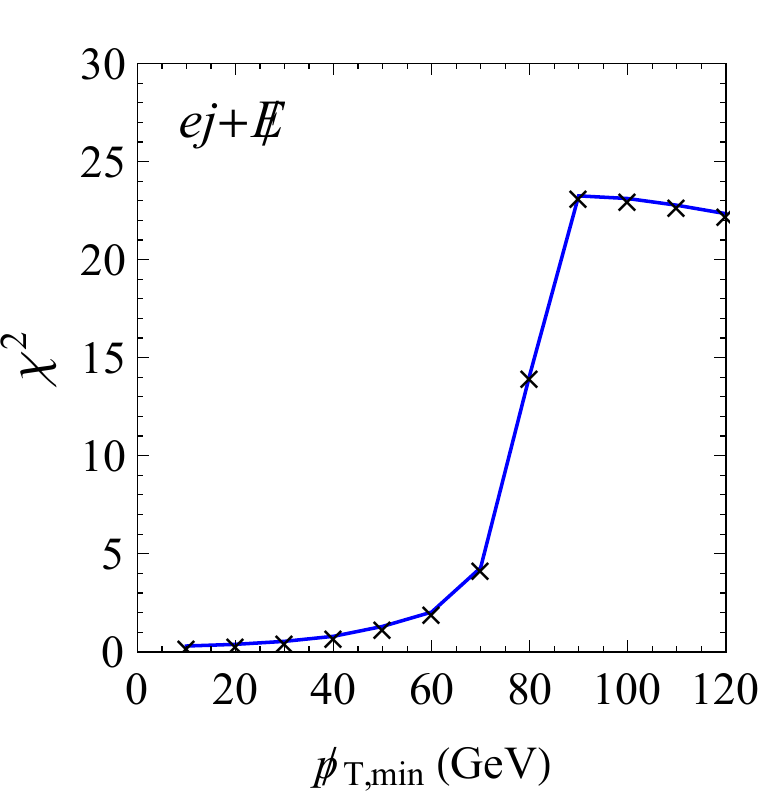}
	\end{center}
	\vspace{-0.3cm}
	\caption{The efficiency of the missing-energy cut for the  $ej + \slashed{E}$ topology. The cut is placed as $\slashed{p}^{}_{\rm T} = |\bm{p}^{}_{\rm T}(e) + \bm{p}^{}_{\rm T}(j)| > \slashed{p}^{}_{\rm T,min}$. }
	\label{fig:SBR}
\end{figure}

\begin{figure*}[t!]
	\begin{center}
		\includegraphics[width=0.45\textwidth]{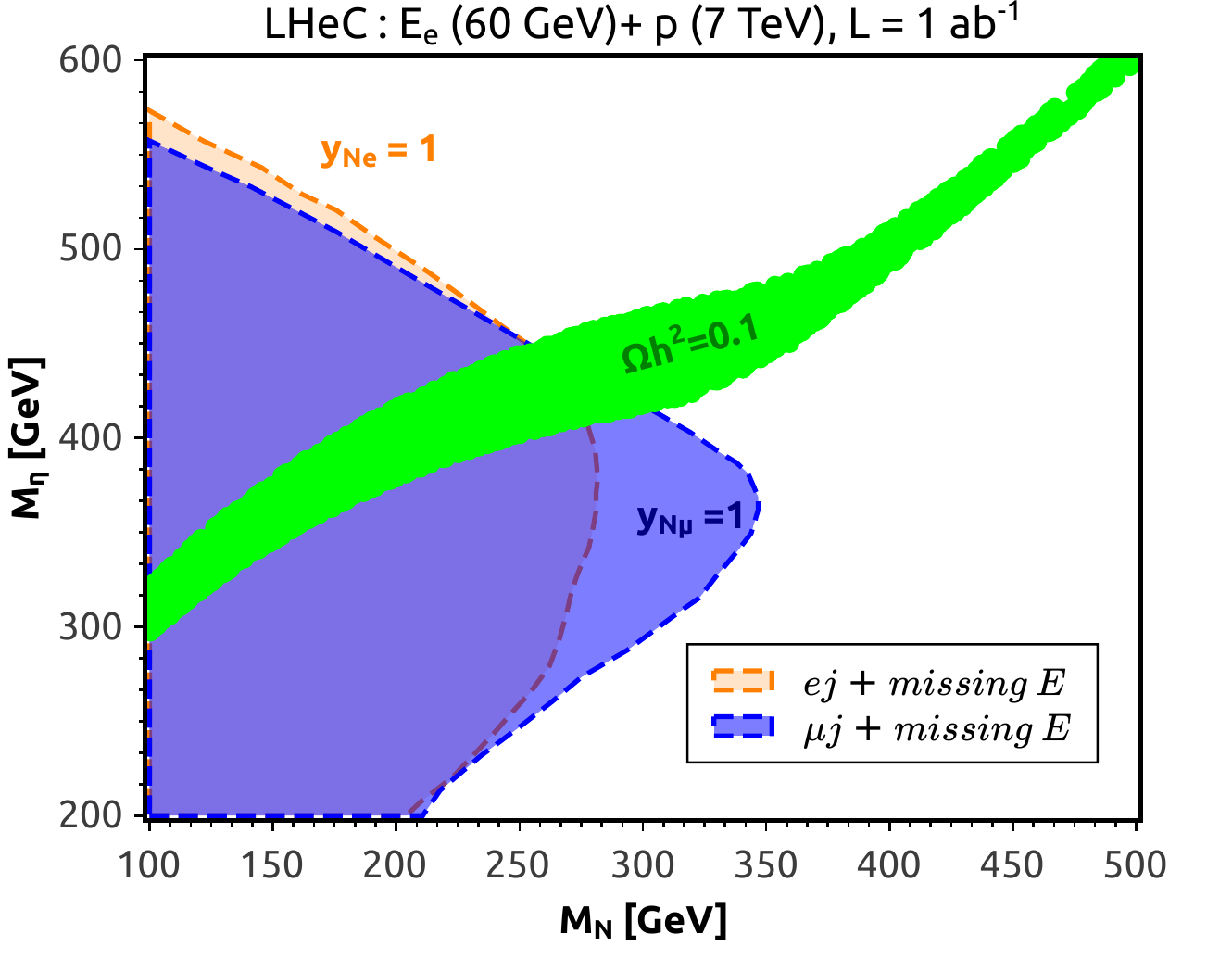}
		\hspace{0.5cm}
		\includegraphics[width=0.45\textwidth]{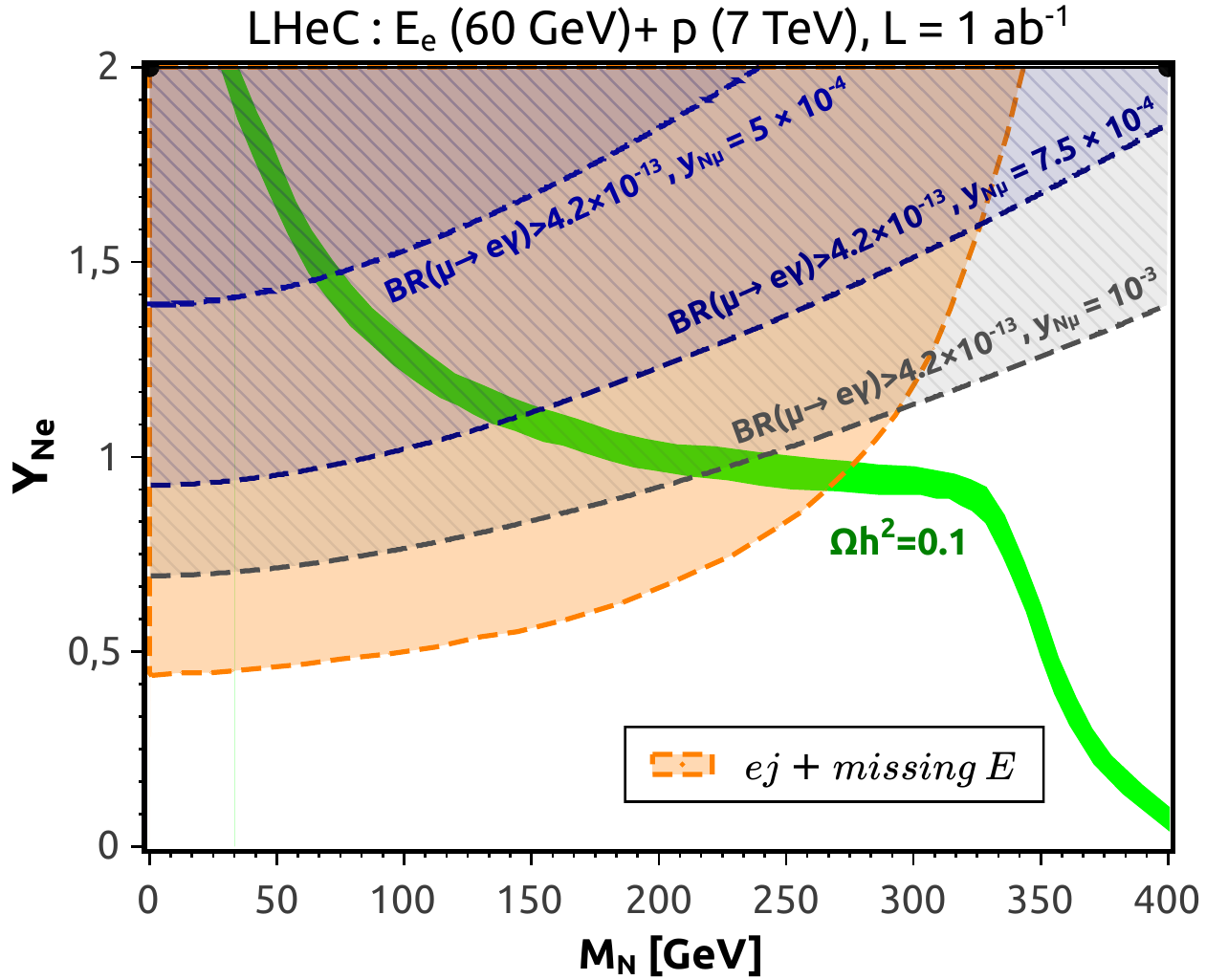}
	\end{center}
	\vspace{-0.3cm}
	\caption{The $3\sigma$ sensitivity region for the LHeC setup based on  60 GeV electrons colliding with  7 TeV protons with an integrated luminosity of $L=1~{\rm ab}^{-1}$. In the left panel, we show the sensitivity to the inert scalar mass $M_\eta$ and Majorana fermion mass $M^{}_{N}$ by fixing the Yukawa coupling $y_{Ne},y_{N\mu}=1$. In the right panel, we fix the inert scalar mass $M_\eta=400~{\rm GeV}$ instead, and scan over $y_{Ne}$ and $M^{}_{N}$. The sensitivities for the $ej + \slashed{E}$ and $\mu j + \slashed{E}$ topologies are displayed with orange and blue regions, respectively. See text for details.
	}
	\label{fig:sens}
\end{figure*}

For the topology $ej + \slashed{E}$, we further impose a cut on $\slashed{p}^{}_{\rm T} = |\bm{p}^{}_{\rm T}(e) + \bm{p}^{}_{\rm T}(j)|$ to reject the $t$-channel backgrounds from direct two-body scattering $eq \to e j$ (fully), $eq \to e j \gamma$ (partly) and $ep \to ejXX$ (partly). The process $eq \to e j \gamma$ contains the initial state radiation of collinear photons and can be efficiently reduced by imposing the missing energy cut. The effectiveness of the cut is shown in Fig.~\ref{fig:SBR} in terms of $\chi^2$ defined in Eq.~(\ref{eq:chi2}), where the model parameters have been taken as $M^{}_{N}=100~{\rm GeV}$, $M_\eta=300~{\rm GeV}$ and $y_{Ne} = 1$. We highlight that $y_{Ne}$ is the Yukawa couplings between the electron and the fermion $N$, whereas $y_{N\mu}$ the Yukawa coupling between the muon and the fermion $N$. Notice, there is only one Majorana fermion $N$ in our collider study, and that for the $\mu j + \slashed{E}$ signal to exist, $y_{Ne}$ ought to be non-zero (see Fig.\ \ref{fig:FD}). The $e j + \slashed{E}$ signal is insensitive to the presence of couplings to muons, however.

That said, we find that a cut $\slashed{p}^{}_{\rm T} \gtrsim 100~{\rm GeV} $ can nearly optimize the signal-to-background ratio of $ej + \slashed{E}$, for which the backgrounds enhanced by collinear processes are almost vanishing. After the cut, the dominant background stems from the process $ep \to ej \nu\nu$. The cut adopted on $\slashed{p}_T$ is based on the $\chi^2$ study (see Fig.\ \ref{fig:SBR}). For $\mu j + \slashed{E}$, we find that the leading background is from the irreducible process $e q \to \mu q \nu^{}_{e} \overline{\nu}^{}_{\mu} $.
The obtained background event numbers for two major signal channels are collected in Table~\ref{table:xsec}.
Before the missing energy cut, the cross sections for $ej + \slashed{E}$ and $\mu j + \slashed{E}$ are almost identical, as their contributing diagrams are the same. The backgrounds from $e j X$ and $e j  X X$ are overwhelmingly large due to the collinear enhancement and can be almost completely removed by imposing the cut $\slashed{p}^{}_{\rm T} \gtrsim 100~{\rm GeV}$.
LHeC's sensitivity to the model is given in Fig.~\ref{fig:sens}. The sensitivity is obtained for the $3\sigma$ confidence level with $\chi^2 = 11.83$ for two degrees of freedom. In the left panel, we fix the Yukawa coupling to one, and then vary the  masses of the inert scalar, $M_\eta$, and Majorana fermion $M^{}_{N}$. In the right panel, we fix the inert scalar mass to be $M_\eta = 400~{\rm GeV}$, and scan over the Yukawa coupling $y^{}_{N}$ and $M^{}_{N}$.  The orange and blue regions correspond to $ej + \slashed{E}$ and $\mu j + \slashed{E}$ events, respectively.
For the $ej + \slashed{E}$ event topology, the cut $\slashed{p}^{}_{\rm T} \gtrsim 100~{\rm GeV} $ has been imposed, while for $\mu j + \slashed{E}$ there is no necessity for such a cut, as explained above. 
We observe that comparable sensitivities can be achieved for $e j + \slashed{E}$ (with $\slashed{p}^{}_{\rm T}$ cut) and $\mu j + \slashed{E}$ events, whereas for some parameter space, one may have a better sensitivity than the other. 
We have investigated each scenario at a time. In other words, for the $ej + \slashed{E}$ study $N$ couples only to the first generation, with $y_{Ne}=1$. Later we probed the $\mu j + \slashed{E}$  signal which requires couplings to both first and second generation of leptons, and we assumed  $y_{Ne}=y_{N\mu}=1$. 
\section{Dark Matter Abundance}
We emphasize that our entire study is based on three quantities, the mass of the dark matter particle, $M_N$,  inert doublet mass, $M_\eta$, and the corresponding Yukawa coupling in Eq.\ (\ref{eq:1}). We remind the reader that we consider each lepton generation at a time. Nevertheless, as far as the dark matter relic density is concerned, they yield the same result. The dark matter relic density is computed under the standard freeze-out regime, where the dark matter decoupling occurs in an epoch in which the Universe is radiation-dominated, via the Boltzmann equation
\begin{equation}
\frac{dn_N}{dt} +3 H n_N =-\langle \sigma v \rangle (n^2_N - n^2_{N,\rm eq})\,.     
\end{equation}
Here $n_N$ is the number density, $H$ the Hubble rate and $\langle \sigma v \rangle$ the annihilation cross section.
The abundance is driven by $t$-channel annihilation processes into leptons mediated by the inert doublet, $\eta$, as dictated by Eq.\ (\ref{eq:1}). We have solved the Boltzmann equation numerically with the \texttt{MicrOMEGAs} package \cite{Belanger:2006is}, and used it to plot the region of parameter space that reproduces the right relic density (green curve) in Fig.\ \ref{fig:sens}. In the left panel, we take $y_{Ne}=1$ and $y_{N\mu}=1$ at a time, and perform a scan in the masses. In the right panel, we take $M_\eta=400$~GeV, and scan over $y_{Ne}$. We assumed $y_{N\mu}$ to be sufficiently small such that the process involving muon pairs is irrelevant for the relic density. The reason for taking a suppressed value for $y_{N\mu}$ has to do with lepton flavor violation probes, but we will address this later. Anyway, looking at Fig.\ \ref{fig:sens}, one can clearly conclude that LHeC can detect a weak scale dark matter candidate that reproduces the right relic density with masses up to $350$ GeV. The relic density curve is subject to shifts if one embeds Eq.\ (\ref{eq:1}) in a non-standard cosmology scenario \cite{Allahverdi:2020bys}, allowing one to get the right relic density for larger dark matter and inert scalar masses.
\begin{figure}[h]
    \centering
    \includegraphics[scale=0.5]{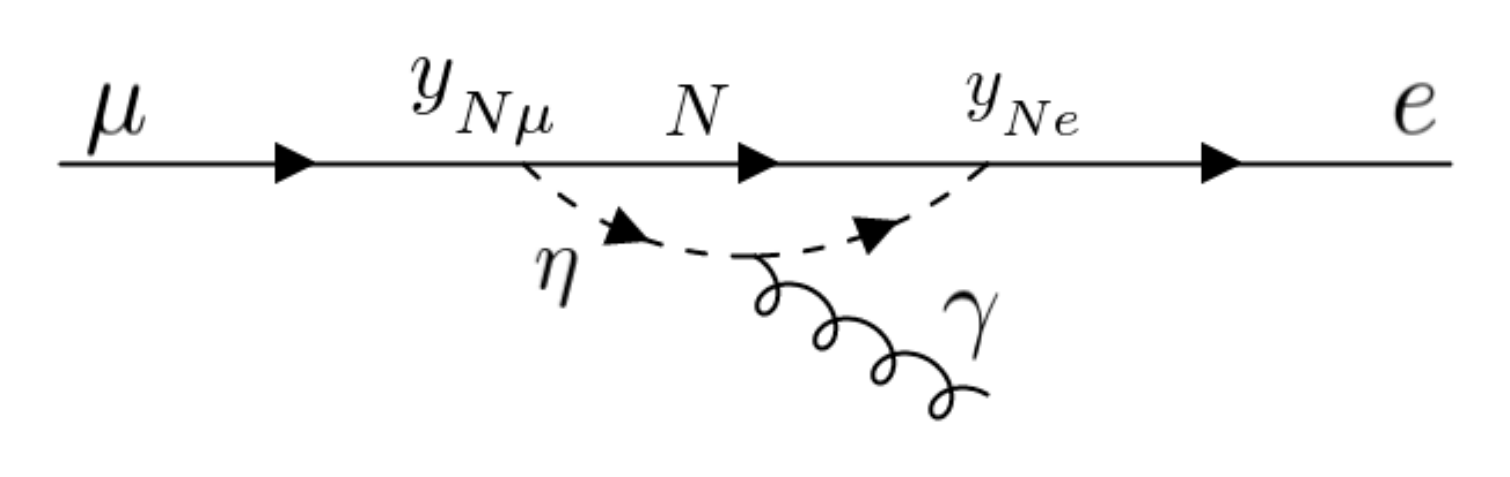}
    \caption{Feynman diagram that leads to $\mu \rightarrow e \gamma$ in our model, when both $y_{N\mu}$ and $y_{Ne}$ are different from zero. See text for details.}
    \label{fig:muontoe}
\end{figure}
\section{Lepton Flavor Violation}
Thus far, our study was based on the presence of one Majorana fermion $N$ that couples either to first or to second generation leptons. It is clear from Fig.\ \ref{fig:sens} that LHeC stands an excellent probe for leptophilic dark matter.  In the first setup, where only couplings to the first generation is present, there is no lepton flavor violation signal, but in the second case where both couplings to first and second generations are included, lepton flavor violation naturally occurs \cite{Adulpravitchai:2009gi,Kubo:2006yx}, as we will explain below.  
When $y_{N\mu}$ and $y_{Ne}$ are non-zero as assumed in the second setup, the lepton flavor violation muon decay occurs are displayed in Fig.\ \ref{fig:muontoe}. The full analytical result for the branching ratio is lengthy, and it relies on the product $(y_{N\mu }\times y_{N e})^2$. This product cannot be of order one, otherwise $BR(\mu \to e\gamma) \sim 10^{-7}$, which is orders of magnitude above the experimental limit $BR(\mu\rightarrow e\gamma)< 4.2 \times 10^{-13}$ \cite{MEG:2016leq}. In the left-panel of Fig.\ \ref{fig:sens} we have shown that one can observe the $\mu j +\slashed{E}$ at the LHeC for larger dark matter masses compared with the $ej +\slashed{E}$ channel.  Now comes the importance of complementary probes. Putting our collider findings into perspective with lepton flavor violation, we can fiercely exclude the entire $\mu j+ \slashed{E}$ region in Fig.\ \ref{fig:sens} because there we had assumed $y_{N \mu}=y_{Ne}=1$.  Lowering $y_{N \mu}$ to values consistent with $\mu\rightarrow e\gamma$ would mean peter out the $\mu j \slashed{E}$ signal.  
Focusing on the $ej +\slashed{E}$ signal, we can estimate what should be the values of $y_{N\mu}$ could yield a $BR(\mu\rightarrow e\gamma)$ consistent with the current limit, while reproducing the correct dark matter relic density. With this in mind, we display in Fig.\ \ref{fig:sens} the $3\sigma$ signal region in $M_N$--$y_{Ne}$ plane, and overlaid contours that yield a $BR(\mu\rightarrow e\gamma)$ in agreement with current bounds for three different Yukawa couplings, namely $y_{N\mu}=10^{-3}$, $y_{N\mu}=7.5\times10^{-4}$ and $y_{N\mu}=5\times10^{-4}$. The overlapping region between the green curve (dark matter relic density) and orange curve (LHeC signal) which is below the contours represents a region where LHeC can detect a weak scale dark matter candidate that reproduces the correct relic density, and could leave some imprints on $\mu\rightarrow e\gamma$ probes in the next generation of experiments, for different coupling strengths to muons.
\section{Direct Detection}
We have checked that direct detection bounds in this simplified case arise at one loop via $Z$-exchange, but at with very suppressed strength. Therefore, we do not expect any direct detection signal from Eq.\ (\ref{eq:1}).  Considering an extended scenario where an additional neutral fermion $N_2$ is present, and a potentially sizeable direct detection signal stems from dark matter-nucleus 
scattering\footnote{Electron recoils are more relevant for light dark matter \cite{Kopp:2009et,Altmannshofer:2016jzy}.}. At  one-loop level, it leads to the effective Lagrangian, 
\begin{equation}
\mathcal{L}_{eff} \supset \alpha \bar{N}_2 \gamma^\mu N \partial^\nu F_{\mu\nu} +  \beta \bar{N}_2  \sigma^{ \mu \nu} N F_{\mu\nu} + \gamma \bar{N}_2 \gamma^\mu N A_\mu,\nonumber\\
\end{equation}where $\alpha$, $\beta$, and $\gamma$ are coefficients that depend on the product $(y_{N \ell} \times y_{N_2 \ell})$, which are the Yukawa couplings of the Majorana fermions with leptons. The terms proportional to coefficients $\alpha$ and $\gamma$ give rise to a charge-charge operator, $\beta$ to a dipole-charge and dipole-dipole operators \cite{Schmidt:2012yg}. Anyway, any direct detection signal is limited by $v_{\rm min}$, which is the minimum velocity at which a dark matter particle can produce a measurable energy recoil,
\begin{eqnarray}
v_{\rm min} =\left(\frac{M_A E_R }{\mu_{AN}}+\delta \right) \frac{1}{\sqrt{2 M_A E_R}},
\end{eqnarray} where $\delta=M_{N_2}-M_{N}$, and $\mu$ is the dark matter-nucleus reduced mass. As we increase the mass difference between these two fermions, we can only induce dark matter-nucleus scattering for large values of $v_{\rm min}$, i.e.\ at the tail of the Maxwell-Boltzmann distribution \cite{Butsky:2015pya}, thus weakening the direct detection bounds. For concreteness, as long as $\delta > 80$~keV, the bounds from direct detection are relatively weak \cite{Schmidt:2012yg}, thus not shown in Fig.\ \ref{fig:sens}. Anyway, in our collider reach study, we have not made any assumption regarding the fermion $N_2$. Therefore, we take it to be sufficiently heavier than $N$, with no prejudice. In summary, direct detection yields less restrictive bounds compared to lepton flavor violation, even in this extended scenario with an additional SM singlet. 
\section{LHC Bounds}
There are constraints on this scenario from LHC, which arise from slepton searches. Since the inert scalar, $\eta$ is a $SU(2)_L$ doublet, one can use the limits from left-handed selectron and smuon searches;  $\eta$ will be pair-produced and then decays back to dark matter. This analysis has been carried out by the ATLAS collaboration \cite{ATLAS:2019lng}, and they could probe inert doublet masses up to $170$~GeV only. That is why we have not included those limits in Fig.\ \ref{fig:sens}. 
\section{Conclusions}
We have shown that LHeC, a collider that features a trade-off in the high energy and intensity frontier, can probe a weak scale fermion dark matter particle, $N$, for masses up to $350$~GeV, through a generic Lagrangian containing an inert doublet. We found that the $ej+\slashed{E}$ and $\mu j + \slashed{E}$ signal topologies are promising and derived $3\sigma$ signal contours for a LHeC setup consisting of  $60$~GeV electrons colliding with $7$~TeV protons and an integrated luminosity of $\mathcal{L}=1$ ab$^{-1}$. We computed the dark matter relic density in the standard freeze-out regime, found that direct detection and LHC bounds are subdominant, and put our findings into perspective with the lepton flavor violation to conclude that the $ej+\slashed{E}$ channels offers a unique opportunity to detect a weak scale dark matter candidate that reproduces the correct relic density and yield a positive signal at $\mu \rightarrow e\gamma$ probes in the next generation of experiments.
\section*{Acknowledgements} 
	\noindent
	The authors thank Avelino Vicente, Carlos Yaguna, and Diego Restrepo for discussions. We thank FAPESP grant 2021/01089-1,
ICTP-SAIFR FAPESP grant 2016/01343-7, CNPq grant 408295/2021-0, Serrapilheira Foundation grant Serra-1912–31613, FONDECYT Grant 1191103 and ANID-Programa Milenio-code $ICN2019\_044$.
GYH is supported by the Alexander von Humboldt Foundation.

\bibliography{reference}

\begin{thebibliography}{21}%
\makeatletter
\providecommand \@ifxundefined [1]{%
 \@ifx{#1\undefined}
}%
\providecommand \@ifnum [1]{%
 \ifnum #1\expandafter \@firstoftwo
 \else \expandafter \@secondoftwo
 \fi
}%
\providecommand \@ifx [1]{%
 \ifx #1\expandafter \@firstoftwo
 \else \expandafter \@secondoftwo
 \fi
}%
\providecommand \natexlab [1]{#1}%
\providecommand \enquote  [1]{``#1''}%
\providecommand \bibnamefont  [1]{#1}%
\providecommand \bibfnamefont [1]{#1}%
\providecommand \citenamefont [1]{#1}%
\providecommand \href@noop [0]{\@secondoftwo}%
\providecommand \href [0]{\begingroup \@sanitize@url \@href}%
\providecommand \@href[1]{\@@startlink{#1}\@@href}%
\providecommand \@@href[1]{\endgroup#1\@@endlink}%
\providecommand \@sanitize@url [0]{\catcode `\\12\catcode `\$12\catcode
  `\&12\catcode `\#12\catcode `\^12\catcode `\_12\catcode `\%12\relax}%
\providecommand \@@startlink[1]{}%
\providecommand \@@endlink[0]{}%
\providecommand \url  [0]{\begingroup\@sanitize@url \@url }%
\providecommand \@url [1]{\endgroup\@href {#1}{\urlprefix }}%
\providecommand \urlprefix  [0]{URL }%
\providecommand \Eprint [0]{\href }%
\providecommand \doibase [0]{http://dx.doi.org/}%
\providecommand \selectlanguage [0]{\@gobble}%
\providecommand \bibinfo  [0]{\@secondoftwo}%
\providecommand \bibfield  [0]{\@secondoftwo}%
\providecommand \translation [1]{[#1]}%
\providecommand \BibitemOpen [0]{}%
\providecommand \bibitemStop [0]{}%
\providecommand \bibitemNoStop [0]{.\EOS\space}%
\providecommand \EOS [0]{\spacefactor3000\relax}%
\providecommand \BibitemShut  [1]{\csname bibitem#1\endcsname}%
\let\auto@bib@innerbib\@empty
\bibitem [{\citenamefont {Abercrombie}\ \emph {et~al.}(2020)\citenamefont
  {Abercrombie} \emph {et~al.}}]{Abercrombie:2015wmb}%
  \BibitemOpen
  \bibfield  {author} {\bibinfo {author} {\bibfnamefont {D.}~\bibnamefont
  {Abercrombie}} \emph {et~al.},\ }\href {\doibase 10.1016/j.dark.2019.100371}
  {\bibfield  {journal} {\bibinfo  {journal} {Phys. Dark Univ.}\ }\textbf
  {\bibinfo {volume} {27}},\ \bibinfo {pages} {100371} (\bibinfo {year}
  {2020})},\ \Eprint {http://arxiv.org/abs/1507.00966} {arXiv:1507.00966
  [hep-ex]} \BibitemShut {NoStop}%
\bibitem [{\citenamefont {Dreiner}\ \emph {et~al.}(2013)\citenamefont
  {Dreiner}, \citenamefont {Huck}, \citenamefont {Kr\"amer}, \citenamefont
  {Schmeier},\ and\ \citenamefont {Tattersall}}]{Dreiner:2012xm}%
  \BibitemOpen
  \bibfield  {author} {\bibinfo {author} {\bibfnamefont {H.}~\bibnamefont
  {Dreiner}}, \bibinfo {author} {\bibfnamefont {M.}~\bibnamefont {Huck}},
  \bibinfo {author} {\bibfnamefont {M.}~\bibnamefont {Kr\"amer}}, \bibinfo
  {author} {\bibfnamefont {D.}~\bibnamefont {Schmeier}}, \ and\ \bibinfo
  {author} {\bibfnamefont {J.}~\bibnamefont {Tattersall}},\ }\href {\doibase
  10.1103/PhysRevD.87.075015} {\bibfield  {journal} {\bibinfo  {journal} {Phys.
  Rev. D}\ }\textbf {\bibinfo {volume} {87}},\ \bibinfo {pages} {075015}
  (\bibinfo {year} {2013})},\ \Eprint {http://arxiv.org/abs/1211.2254}
  {arXiv:1211.2254 [hep-ph]} \BibitemShut {NoStop}%
\bibitem [{\citenamefont {Abelleira~Fernandez}\ \emph
  {et~al.}(2012)\citenamefont {Abelleira~Fernandez} \emph
  {et~al.}}]{LHeCStudyGroup:2012zhm}%
  \BibitemOpen
  \bibfield  {author} {\bibinfo {author} {\bibfnamefont {J.~L.}\ \bibnamefont
  {Abelleira~Fernandez}} \emph {et~al.} (\bibinfo {collaboration} {LHeC Study
  Group}),\ }\href {\doibase 10.1088/0954-3899/39/7/075001} {\bibfield
  {journal} {\bibinfo  {journal} {J. Phys. G}\ }\textbf {\bibinfo {volume}
  {39}},\ \bibinfo {pages} {075001} (\bibinfo {year} {2012})},\ \Eprint
  {http://arxiv.org/abs/1206.2913} {arXiv:1206.2913 [physics.acc-ph]}
  \BibitemShut {NoStop}%
\bibitem [{\citenamefont {Abada}\ \emph {et~al.}(2019)\citenamefont {Abada}
  \emph {et~al.}}]{FCC:2018evy}%
  \BibitemOpen
  \bibfield  {author} {\bibinfo {author} {\bibfnamefont {A.}~\bibnamefont
  {Abada}} \emph {et~al.} (\bibinfo {collaboration} {FCC}),\ }\href {\doibase
  10.1140/epjst/e2019-900045-4} {\bibfield  {journal} {\bibinfo  {journal}
  {Eur. Phys. J. ST}\ }\textbf {\bibinfo {volume} {228}},\ \bibinfo {pages}
  {261} (\bibinfo {year} {2019})}\BibitemShut {NoStop}%
\bibitem [{\citenamefont {Baer}\ \emph {et~al.}(2013)\citenamefont {Baer} \emph
  {et~al.}}]{Baer:2013cma}%
  \BibitemOpen
  \bibfield  {author} {\bibinfo {author} {\bibfnamefont {H.}~\bibnamefont
  {Baer}} \emph {et~al.},\ }\href@noop {} {\  (\bibinfo {year} {2013})},\
  \Eprint {http://arxiv.org/abs/1306.6352} {arXiv:1306.6352 [hep-ph]}
  \BibitemShut {NoStop}%
\bibitem [{\citenamefont {Ma}(2006)}]{Ma:2006km}%
  \BibitemOpen
  \bibfield  {author} {\bibinfo {author} {\bibfnamefont {E.}~\bibnamefont
  {Ma}},\ }\href {\doibase 10.1103/PhysRevD.73.077301} {\bibfield  {journal}
  {\bibinfo  {journal} {Phys. Rev. D}\ }\textbf {\bibinfo {volume} {73}},\
  \bibinfo {pages} {077301} (\bibinfo {year} {2006})},\ \Eprint
  {http://arxiv.org/abs/hep-ph/0601225} {arXiv:hep-ph/0601225} \BibitemShut
  {NoStop}%
\bibitem [{\citenamefont {Schmidt}\ \emph {et~al.}(2012)\citenamefont
  {Schmidt}, \citenamefont {Schwetz},\ and\ \citenamefont
  {Toma}}]{Schmidt:2012yg}%
  \BibitemOpen
  \bibfield  {author} {\bibinfo {author} {\bibfnamefont {D.}~\bibnamefont
  {Schmidt}}, \bibinfo {author} {\bibfnamefont {T.}~\bibnamefont {Schwetz}}, \
  and\ \bibinfo {author} {\bibfnamefont {T.}~\bibnamefont {Toma}},\ }\href
  {\doibase 10.1103/PhysRevD.85.073009} {\bibfield  {journal} {\bibinfo
  {journal} {Phys. Rev. D}\ }\textbf {\bibinfo {volume} {85}},\ \bibinfo
  {pages} {073009} (\bibinfo {year} {2012})},\ \Eprint
  {http://arxiv.org/abs/1201.0906} {arXiv:1201.0906 [hep-ph]} \BibitemShut
  {NoStop}%
\bibitem [{\citenamefont {Toma}\ and\ \citenamefont
  {Vicente}(2014)}]{Toma:2013zsa}%
  \BibitemOpen
  \bibfield  {author} {\bibinfo {author} {\bibfnamefont {T.}~\bibnamefont
  {Toma}}\ and\ \bibinfo {author} {\bibfnamefont {A.}~\bibnamefont {Vicente}},\
  }\href {\doibase 10.1007/JHEP01(2014)160} {\bibfield  {journal} {\bibinfo
  {journal} {JHEP}\ }\textbf {\bibinfo {volume} {01}},\ \bibinfo {pages} {160}
  (\bibinfo {year} {2014})},\ \Eprint {http://arxiv.org/abs/1312.2840}
  {arXiv:1312.2840 [hep-ph]} \BibitemShut {NoStop}%
\bibitem [{\citenamefont {Lindner}\ \emph {et~al.}(2018)\citenamefont
  {Lindner}, \citenamefont {Platscher},\ and\ \citenamefont
  {Queiroz}}]{Lindner:2016bgg}%
  \BibitemOpen
  \bibfield  {author} {\bibinfo {author} {\bibfnamefont {M.}~\bibnamefont
  {Lindner}}, \bibinfo {author} {\bibfnamefont {M.}~\bibnamefont {Platscher}},
  \ and\ \bibinfo {author} {\bibfnamefont {F.~S.}\ \bibnamefont {Queiroz}},\
  }\href {\doibase 10.1016/j.physrep.2017.12.001} {\bibfield  {journal}
  {\bibinfo  {journal} {Phys. Rept.}\ }\textbf {\bibinfo {volume} {731}},\
  \bibinfo {pages} {1} (\bibinfo {year} {2018})},\ \Eprint
  {http://arxiv.org/abs/1610.06587} {arXiv:1610.06587 [hep-ph]} \BibitemShut
  {NoStop}%
\bibitem [{\citenamefont {Cottin}\ \emph {et~al.}(2021)\citenamefont {Cottin},
  \citenamefont {Fischer}, \citenamefont {Mandal}, \citenamefont {Mitra},\ and\
  \citenamefont {Padhan}}]{Cottin:2021tfo}%
  \BibitemOpen
  \bibfield  {author} {\bibinfo {author} {\bibfnamefont {G.}~\bibnamefont
  {Cottin}}, \bibinfo {author} {\bibfnamefont {O.}~\bibnamefont {Fischer}},
  \bibinfo {author} {\bibfnamefont {S.}~\bibnamefont {Mandal}}, \bibinfo
  {author} {\bibfnamefont {M.}~\bibnamefont {Mitra}}, \ and\ \bibinfo {author}
  {\bibfnamefont {R.}~\bibnamefont {Padhan}},\ }\href@noop {} {\  (\bibinfo
  {year} {2021})},\ \Eprint {http://arxiv.org/abs/2104.13578} {arXiv:2104.13578
  [hep-ph]} \BibitemShut {NoStop}%
\bibitem [{\citenamefont {Lindner}\ \emph {et~al.}(2016)\citenamefont
  {Lindner}, \citenamefont {Queiroz}, \citenamefont {Rodejohann},\ and\
  \citenamefont {Yaguna}}]{Lindner:2016lxq}%
  \BibitemOpen
  \bibfield  {author} {\bibinfo {author} {\bibfnamefont {M.}~\bibnamefont
  {Lindner}}, \bibinfo {author} {\bibfnamefont {F.~S.}\ \bibnamefont
  {Queiroz}}, \bibinfo {author} {\bibfnamefont {W.}~\bibnamefont {Rodejohann}},
  \ and\ \bibinfo {author} {\bibfnamefont {C.~E.}\ \bibnamefont {Yaguna}},\
  }\href {\doibase 10.1007/JHEP06(2016)140} {\bibfield  {journal} {\bibinfo
  {journal} {JHEP}\ }\textbf {\bibinfo {volume} {06}},\ \bibinfo {pages} {140}
  (\bibinfo {year} {2016})},\ \Eprint {http://arxiv.org/abs/1604.08596}
  {arXiv:1604.08596 [hep-ph]} \BibitemShut {NoStop}%
\bibitem [{\citenamefont {Belanger}\ \emph {et~al.}(2007)\citenamefont
  {Belanger}, \citenamefont {Boudjema}, \citenamefont {Pukhov},\ and\
  \citenamefont {Semenov}}]{Belanger:2006is}%
  \BibitemOpen
  \bibfield  {author} {\bibinfo {author} {\bibfnamefont {G.}~\bibnamefont
  {Belanger}}, \bibinfo {author} {\bibfnamefont {F.}~\bibnamefont {Boudjema}},
  \bibinfo {author} {\bibfnamefont {A.}~\bibnamefont {Pukhov}}, \ and\ \bibinfo
  {author} {\bibfnamefont {A.}~\bibnamefont {Semenov}},\ }\href {\doibase
  10.1016/j.cpc.2006.11.008} {\bibfield  {journal} {\bibinfo  {journal}
  {Comput. Phys. Commun.}\ }\textbf {\bibinfo {volume} {176}},\ \bibinfo
  {pages} {367} (\bibinfo {year} {2007})},\ \Eprint
  {http://arxiv.org/abs/hep-ph/0607059} {arXiv:hep-ph/0607059} \BibitemShut
  {NoStop}%
\bibitem [{\citenamefont {Allahverdi}\ \emph {et~al.}(2020)\citenamefont
  {Allahverdi} \emph {et~al.}}]{Allahverdi:2020bys}%
  \BibitemOpen
  \bibfield  {author} {\bibinfo {author} {\bibfnamefont {R.}~\bibnamefont
  {Allahverdi}} \emph {et~al.},\ }\href {\doibase 10.21105/astro.2006.16182} {\
   (\bibinfo {year} {2020}),\ 10.21105/astro.2006.16182},\ \Eprint
  {http://arxiv.org/abs/2006.16182} {arXiv:2006.16182 [astro-ph.CO]}
  \BibitemShut {NoStop}%
\bibitem [{\citenamefont {Adulpravitchai}\ \emph {et~al.}(2009)\citenamefont
  {Adulpravitchai}, \citenamefont {Lindner},\ and\ \citenamefont
  {Merle}}]{Adulpravitchai:2009gi}%
  \BibitemOpen
  \bibfield  {author} {\bibinfo {author} {\bibfnamefont {A.}~\bibnamefont
  {Adulpravitchai}}, \bibinfo {author} {\bibfnamefont {M.}~\bibnamefont
  {Lindner}}, \ and\ \bibinfo {author} {\bibfnamefont {A.}~\bibnamefont
  {Merle}},\ }\href {\doibase 10.1103/PhysRevD.80.055031} {\bibfield  {journal}
  {\bibinfo  {journal} {Phys. Rev. D}\ }\textbf {\bibinfo {volume} {80}},\
  \bibinfo {pages} {055031} (\bibinfo {year} {2009})},\ \Eprint
  {http://arxiv.org/abs/0907.2147} {arXiv:0907.2147 [hep-ph]} \BibitemShut
  {NoStop}%
\bibitem [{\citenamefont {Kubo}\ \emph {et~al.}(2006)\citenamefont {Kubo},
  \citenamefont {Ma},\ and\ \citenamefont {Suematsu}}]{Kubo:2006yx}%
  \BibitemOpen
  \bibfield  {author} {\bibinfo {author} {\bibfnamefont {J.}~\bibnamefont
  {Kubo}}, \bibinfo {author} {\bibfnamefont {E.}~\bibnamefont {Ma}}, \ and\
  \bibinfo {author} {\bibfnamefont {D.}~\bibnamefont {Suematsu}},\ }\href
  {\doibase 10.1016/j.physletb.2006.08.085} {\bibfield  {journal} {\bibinfo
  {journal} {Phys. Lett. B}\ }\textbf {\bibinfo {volume} {642}},\ \bibinfo
  {pages} {18} (\bibinfo {year} {2006})},\ \Eprint
  {http://arxiv.org/abs/hep-ph/0604114} {arXiv:hep-ph/0604114} \BibitemShut
  {NoStop}%
\bibitem [{\citenamefont {Baldini}\ \emph {et~al.}(2016)\citenamefont {Baldini}
  \emph {et~al.}}]{MEG:2016leq}%
  \BibitemOpen
  \bibfield  {author} {\bibinfo {author} {\bibfnamefont {A.~M.}\ \bibnamefont
  {Baldini}} \emph {et~al.} (\bibinfo {collaboration} {MEG}),\ }\href {\doibase
  10.1140/epjc/s10052-016-4271-x} {\bibfield  {journal} {\bibinfo  {journal}
  {Eur. Phys. J. C}\ }\textbf {\bibinfo {volume} {76}},\ \bibinfo {pages} {434}
  (\bibinfo {year} {2016})},\ \Eprint {http://arxiv.org/abs/1605.05081}
  {arXiv:1605.05081 [hep-ex]} \BibitemShut {NoStop}%
\bibitem [{Note1()}]{Note1}%
  \BibitemOpen
  \bibinfo {note} {Electron recoils are more relevant for light dark matter
  \cite {Kopp:2009et,Altmannshofer:2016jzy}.}\BibitemShut {Stop}%
\bibitem [{\citenamefont {Butsky}\ \emph {et~al.}(2016)\citenamefont {Butsky},
  \citenamefont {Macci\`o}, \citenamefont {Dutton}, \citenamefont {Wang},
  \citenamefont {Obreja}, \citenamefont {Stinson}, \citenamefont {Penzo},
  \citenamefont {Kang}, \citenamefont {Keller},\ and\ \citenamefont
  {Wadsley}}]{Butsky:2015pya}%
  \BibitemOpen
  \bibfield  {author} {\bibinfo {author} {\bibfnamefont {I.}~\bibnamefont
  {Butsky}}, \bibinfo {author} {\bibfnamefont {A.~V.}\ \bibnamefont
  {Macci\`o}}, \bibinfo {author} {\bibfnamefont {A.~A.}\ \bibnamefont
  {Dutton}}, \bibinfo {author} {\bibfnamefont {L.}~\bibnamefont {Wang}},
  \bibinfo {author} {\bibfnamefont {A.}~\bibnamefont {Obreja}}, \bibinfo
  {author} {\bibfnamefont {G.~S.}\ \bibnamefont {Stinson}}, \bibinfo {author}
  {\bibfnamefont {C.}~\bibnamefont {Penzo}}, \bibinfo {author} {\bibfnamefont
  {X.}~\bibnamefont {Kang}}, \bibinfo {author} {\bibfnamefont {B.~W.}\
  \bibnamefont {Keller}}, \ and\ \bibinfo {author} {\bibfnamefont
  {J.}~\bibnamefont {Wadsley}},\ }\href {\doibase 10.1093/mnras/stw1688}
  {\bibfield  {journal} {\bibinfo  {journal} {Mon. Not. Roy. Astron. Soc.}\
  }\textbf {\bibinfo {volume} {462}},\ \bibinfo {pages} {663} (\bibinfo {year}
  {2016})},\ \Eprint {http://arxiv.org/abs/1503.04814} {arXiv:1503.04814
  [astro-ph.GA]} \BibitemShut {NoStop}%
\bibitem [{\citenamefont {Aad}\ \emph {et~al.}(2020)\citenamefont {Aad} \emph
  {et~al.}}]{ATLAS:2019lng}%
  \BibitemOpen
  \bibfield  {author} {\bibinfo {author} {\bibfnamefont {G.}~\bibnamefont
  {Aad}} \emph {et~al.} (\bibinfo {collaboration} {ATLAS}),\ }\href {\doibase
  10.1103/PhysRevD.101.052005} {\bibfield  {journal} {\bibinfo  {journal}
  {Phys. Rev. D}\ }\textbf {\bibinfo {volume} {101}},\ \bibinfo {pages}
  {052005} (\bibinfo {year} {2020})},\ \Eprint
  {http://arxiv.org/abs/1911.12606} {arXiv:1911.12606 [hep-ex]} \BibitemShut
  {NoStop}%
\bibitem [{\citenamefont {Kopp}\ \emph {et~al.}(2009)\citenamefont {Kopp},
  \citenamefont {Niro}, \citenamefont {Schwetz},\ and\ \citenamefont
  {Zupan}}]{Kopp:2009et}%
  \BibitemOpen
  \bibfield  {author} {\bibinfo {author} {\bibfnamefont {J.}~\bibnamefont
  {Kopp}}, \bibinfo {author} {\bibfnamefont {V.}~\bibnamefont {Niro}}, \bibinfo
  {author} {\bibfnamefont {T.}~\bibnamefont {Schwetz}}, \ and\ \bibinfo
  {author} {\bibfnamefont {J.}~\bibnamefont {Zupan}},\ }\href {\doibase
  10.1103/PhysRevD.80.083502} {\bibfield  {journal} {\bibinfo  {journal} {Phys.
  Rev. D}\ }\textbf {\bibinfo {volume} {80}},\ \bibinfo {pages} {083502}
  (\bibinfo {year} {2009})},\ \Eprint {http://arxiv.org/abs/0907.3159}
  {arXiv:0907.3159 [hep-ph]} \BibitemShut {NoStop}%
\bibitem [{\citenamefont {Altmannshofer}\ \emph {et~al.}(2016)\citenamefont
  {Altmannshofer}, \citenamefont {Gori}, \citenamefont {Profumo},\ and\
  \citenamefont {Queiroz}}]{Altmannshofer:2016jzy}%
  \BibitemOpen
  \bibfield  {author} {\bibinfo {author} {\bibfnamefont {W.}~\bibnamefont
  {Altmannshofer}}, \bibinfo {author} {\bibfnamefont {S.}~\bibnamefont {Gori}},
  \bibinfo {author} {\bibfnamefont {S.}~\bibnamefont {Profumo}}, \ and\
  \bibinfo {author} {\bibfnamefont {F.~S.}\ \bibnamefont {Queiroz}},\ }\href
  {\doibase 10.1007/JHEP12(2016)106} {\bibfield  {journal} {\bibinfo  {journal}
  {JHEP}\ }\textbf {\bibinfo {volume} {12}},\ \bibinfo {pages} {106} (\bibinfo
  {year} {2016})},\ \Eprint {http://arxiv.org/abs/1609.04026} {arXiv:1609.04026
  [hep-ph]} \BibitemShut {NoStop}%
\end{thebibliography}%

\end{document}